\documentclass[twocolumn, fleqn]{article}

\usepackage[utf8]{inputenc}
\usepackage[T1]{fontenc}
\usepackage[a4paper, margin=0.75in]{geometry}
\usepackage[hidelinks]{hyperref}
\usepackage{amsmath}
\usepackage[list-units=single]{siunitx}
\usepackage[version=4]{mhchem}
\usepackage{tikz}
\usepackage{graphicx}
\usepackage{float}
\usepackage{subcaption}
\usepackage{booktabs}
\usepackage{multicol}
\usepackage[capitalize, noabbrev, nameinlink]{cleveref}
\usepackage[style=nature]{biblatex}
\usepackage{titling}
\usepackage{authblk}
\usepackage{xcolor}
\usepackage{xr}

\DeclareSIUnit{\angstrom}{\textup{\smash{Å}\vphantom{A}}}
\DeclareSIUnit{\atom}{\text{atom}}
\DeclareSIUnit{\bar}{\text{bar}}

\hypersetup{
    hypertexnames=false
}

\newcommand{\subfiglabel}[1]{\begin{subfigure}{0em}\phantomsubcaption\label{#1}\end{subfigure}}

\newcommand*{\citet}[1]{\textcite{#1}}

\title{Breaking scaling relations with inverse catalysts: a machine learning exploration of trends in \ce{CO2} hydrogenation energy barriers}
\author[1]{Luuk H. E. Kempen}
\author[1]{Marius Juul Nielsen}
\author[1,2,*]{Mie Andersen}
\affil[1]{Center for Interstellar Catalysis, Department of Physics and Astronomy, Aarhus University, Aarhus C, Denmark}
\affil[2]{Aarhus Institute of Advanced Studies, Aarhus University, Aarhus C, Denmark}
\affil[*]{Corresponding author: \url{mie@phys.au.dk}}

\begin{document}

\twocolumn[
    \begin{@twocolumnfalse}
        \maketitle

        The conversion of \ce{CO2} into useful products such as methanol is a key strategy for abating climate change and our dependence on fossil fuels.
Developing new catalysts for this process is costly and time-consuming and can thus benefit from computational exploration of possible active sites.
However, this is complicated by the complexity of the materials and reaction networks.
Here, we present a workflow for exploring transition states of elementary reaction steps at inverse catalysts, which is based on the training of a neural network-based machine learning interatomic potential.
We focus on the crucial formate intermediate and its formation over nanoclusters of indium oxide supported on \ce{Cu(111)}.
The speedup compared to an approach purely based on density functional theory allows us to probe a wide variety of active sites found at nanoclusters of different sizes and stoichiometries.
Analysis of the obtained set of transition state geometries reveals different structure--activity trends at the edge or interior of the nanoclusters.
Furthermore, the identified geometries allow for the breaking of linear scaling relations, which could be a key underlying reason for the excellent catalytic performance of inverse catalysts observed in experiments.

    \end{@twocolumnfalse}

    \vspace*{2em}
]

\section{Introduction}

In industry, the \ce{Cu/ZnO/Al2O3} catalyst is currently employed in the conventional method of methanol production from \ce{CO} and \ce{CO2}~\cite{guillopez2019}.
Recently, \ce{ZnO_x} overlayers have been observed on the \ce{Cu} nanoparticles~\cite{lunkenbein2015}, possibly hinting to metal oxides playing an important role in the catalytic process.
Furthermore, inverse catalysts---metal oxide nanoparticles on a metal support---have been shown to be more active than their respective conventional counterparts for \ce{CO2} hydrogenation~\cite{graciani2014,senanayake2016,wu2020}.
Hence, inverse catalysts are a promising class of catalyst materials.

In computational screening of catalysts, microkinetic modeling is an invaluable technique for predicting catalytic activity~\cite{motagamwala2020}.
Part of constructing such a model is collecting adsorption energies and activation energies of elementary reaction steps on the catalyst surface, which in turn requires knowledge of the surface geometry and active sites.
These inputs are typically obtained from density functional theory (DFT) calculations.
Moreover, for metal surfaces, linear scaling relations often allow for cheap estimation of activation energies from reaction energies through Brønsted--Evans--Polanyi (BEP) relations~\cite{bronsted1928,evans1936,norskov2002,Michaelides2003}.

\begin{figure*}
    \centering
    \includegraphics{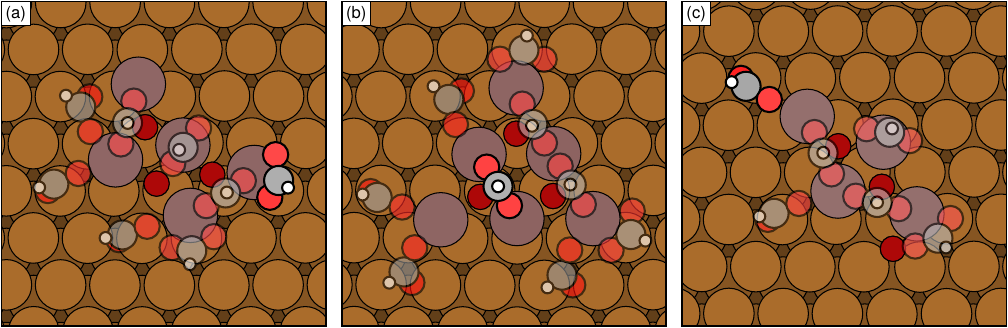}
    \caption{A subset of (unoptimized) possible formate binding sites on three \ce{In_{y}O_{x}/Cu(111)} inverse nanocluster geometries. Only non-overlapping binding sites are shown for ease of visualization. On each geometry, one binding site is highlighted to indicate three different types of binding sites: (a) on-top; (b) bridging between two cluster atoms; (c) bridging at the metal--oxide interface. Color legend: \ce{Cu} (brown), \ce{In} (purple), \ce{O} in oxide (dark red), \ce{O} in formate (light red), \ce{C} (gray), \ce{H} (white).}
    \label{fig:formate-placements}
\end{figure*}

Inverse catalysts are much more difficult to model than low-index metal facets due to the additional complexity in possible compositions and structures~\cite{Reichenbach2019,kempen2025}.
Furthermore, the structures can be highly asymmetric, leading to many possible active sites to consider in the modelling (see \cref{fig:formate-placements}).
And to make matters worse, previous work on oxide-supported metal catalysts have revealed that linear scaling relations may not generally hold at metal--oxide interfaces~\cite{mehta2017,lustemberg2020}, which necessitates the development of other computationally tractable approaches for obtaining activation energies.

Activation energies can be calculated by finding saddle points on the potential energy surface. Common methods for this task are interpolation methods, e.g., the nudged elastic band (NEB) method~\cite{jonsson1998}, that optimize a chain of images along the reaction path, or local methods, e.g., the dimer method~\cite{henkelman1999}, that optimize one or two images near the transition state (TS).
However, all these methods can be hard to converge and often require many potential energy evaluations.
Combined with the immense search space of possible active sites, this renders a pure DFT approach to TS search at inverse catalysts intractible.

In recent years, machine learning interatomic potentials (MLIPs) have become viable for obtaining approximations of the potential energy surface at near-DFT-level accuracy for a fraction of the computational cost.
This unlocks the possibility of high-throughput screening of elementary reaction steps across cluster compositions, structures and active sites.
In this work, we develop a ML-driven workflow for obtaining accurate TS guesses for elementary reaction steps.
The employed MLIPs are based on the Gaussian moment neural network (GM-NN) architecture~\cite{zaverkin2020,zaverkin2021}.
We focus here on the \ce{In_yO_x/Cu(111)} inverse catalyst system, for which stable cluster compositions and structures were identified in our previous work~\cite{kempen2025}.
The rationale for exploring this system in detail is (i) that \ce{Cu} and indium oxide are interesting catalysts for \ce{CO2} conversion in their own right \cite{Grabow2011,frei2018} and (ii) that much less attention has been devoted to this system compared to other similar inverse catalyst systems such as \ce{Zn_yO_x/Cu(111)}~\cite{kattel2017,Reichenbach2018}.
And we focus on the formation of formate (\ce{HCOO}) from \ce{CO2} and \ce{H}, as formate has been found to be a key intermediate in \ce{CO2} hydrogenation to methanol~\cite{kattel2017,frei2018,cheula2024}.

We show that our developed workflow can not only predict accurate TS structures for active sites considered in the training, but also generalizes well to new active site motifs.
This capability allows us to assemble a large data set of TS structures across inverse catalyst compositions and structures at a fraction of the computational cost required for traditional DFT-based approaches.
Analysis of this TS data set reveals siginificant structure-activity trends.
In particular, we show that the active sites at inverse catalysts can break linear scaling relations, which may be a key reason for the high catalytic activity reported in experiments~\cite{graciani2014,senanayake2016,wu2020,Prabhakar2024}.

\section{Methods}

\subsection{Density functional theory calculations}  \label{sec:methods/dft}

All training and validation data for the machine learning models was obtained with density functional theory (DFT).
The DFT calculations were carried out using the GPAW package~\cite{mortensen2005,enkovaara2010,mortensen2024}, which implements the projector augmented-wave method~\cite{blochl1994}.
GPAW was used in the plane-wave mode and interfaced with the atomic simulation environment (ASE) package~\cite{larsen2017}.
The exchange--correlation functional was approximated using the Perdew--Burke--Ernzerhof (PBE) functional~\cite{perdew1996}.
All calculations were further corrected by a DFT-D4 dispersion correction~\cite{caldeweyher2017,caldeweyher2019,caldeweyher2020}.
The metal slab model was not modified from our previous work~\cite{kempen2025}; it was constructed using a \numproduct{7x7x4} supercell with a vacuum of \qty{18}{\angstrom} between the metal layers in the direction perpendicular to the surface.
Orbital occupancies were set using the Fermi--Dirac distribution with a width of \qty{0.10}{\electronvolt}.

To speed up data collection, optimization performed to obtain structures for training data was done with low-level DFT settings: a plane-wave cutoff of \qty{400}{\electronvolt} was employed and the Brillouin zone was sampled using the $\Gamma$ point.
Structure optimization was performed using the Broyden--Fletcher--Goldfarb--Shanno (BFGS) algorithm with an atomic force convergence of $F_\mathrm{max} = \qty{0.05}{\electronvolt\per\angstrom}$, where the topmost layer of the metal surface, the nanocluster, and the adsorbate atoms were allowed to relax.
DFT validation of TSs was performed using a local bond climbing method, beginning from a suitable ML-optimized guess structure, as outlined in \cref{sec:methods/ts-search}.
For this we employed a plane-wave cutoff of \qty{600}{\electronvolt} and $\Gamma$ point sampling of the Brillouin zone.
Structures selected to be part of the training data as well as optimized TS structures were finally refined with a single-point higher-level DFT calculation employing a plane-wave cutoff of \qty{600}{\electronvolt} and a \numproduct{2x2x1} Monkhorst--Pack $k$-point grid~\cite{monkhorst1976} for the Brillouin zone sampling.

Formation energies, $E_\mathrm{f}$, of adsorbates (formate or the TS for its formation) were calculated as
\begin{equation}
    E_\mathrm{f} = E_\mathrm{cluster + ads} - E_\mathrm{cluster} - E_{\ce{CO2 (g)}} - \frac{1}{2} E_{\ce{H2 (g)}}
\end{equation}
where $E_\mathrm{cluster + ads}$ is the energy of the nanocluster with the adsorbate, $E_\mathrm{cluster}$ is the energy of the clean nanocluster, and $E_{\ce{CO2 (g)}}$ and $E_{\ce{H2 (g)}}$ are the energies of gas-phase \ce{CO2} and \ce{H2}, respectively.

\subsection{Machine learning interatomic potential}  \label{sec:methods/gmnn}

In this work, we use the Gaussian moment neural network (GM-NN) architecture~\cite{zaverkin2020,zaverkin2021} as machine learning interatomic potential, as implemented in the Apax package~\cite{schafer2024}.
In this architecture, the total energy of a system $S$ can be decomposed into a sum of atomic energies~\cite{behler2007}
\begin{equation}
    \hat{E}(S) \approx \sum_{i=1}^{N_\mathrm{at}} \hat{E}_i (\mathbf{G}_i),
\end{equation}
where $\mathbf{G}_i$ is the local atomic representation of atom $i$.
This representation is constructed from the pair distance vectors $\mathbf{r}_{ij} = \mathbf{r}_i - \mathbf{r}_j$ for all atomic pairs within a predefined cutoff radius.
These pair distance vectors are used in a translationally and permutationally invariant tensor-valued function, which is converted into features $\mathbf{G}_i$ that are also rotationally invariant through full tensor contractions.
These features are used as input to a fully connected feedforward neural network to give atomic energy and force predictions.

The GM-NNs were generally trained in two steps.
First, a single network was trained using mean squared error (MSE) on energies and forces as loss function, where the force error was weighted by a factor of $3 N_\mathrm{at}$ (three times the number of atoms) relative to the energy error.
With the exception of last-layer weights, the parameters of this network were then transferred to a shallow ensemble with \num{10} output heads for further training with a negative log-likelihood (NLL) loss function.
In this network, the last layer is split into an ensemble of models, following the direct propagation of shallow ensembles (DPOSE) architecture~\cite{kellner2024}.
Each model head makes its own predictions, and the model output is given by the mean of the ensemble predictions.
This also allows for evaluation of an ensemble uncertainty \emph{via} the variance of the individual predictions.

Hyperparameters for the GM-NN were obtained by optimization using the Optuna framework~\cite{akiba2019}.
After setting up a space of hyperparameters for the GM-NN models, Optuna dynamically selects suggestions for hyperparameter values based on earlier results.
With these suggestions, a GM-NN model is trained and the resulting loss value on validation data is used to feed hyperparameter choices for the next trials.
The set of hyperparameters resulting in the optimal validation loss was then chosen for further GM-NN model training.
Learning rates were reduced for training the shallow ensemble.
An overview of the optimized hyperparameters for the used networks is given in \cref{supp/tab:model-hyperparameters}.

In total, \num{10} separate shallow-ensemble GM-NN models were trained fully independently, with the goal of obtaining models to individually predict slightly different TS geometries.

\subsection{Training data acquisition}  \label{sec:methods/training-data}

To train the GM-NN models for TS prediction, a data set containing structures relevant for the reaction pathway was constructed.
Here we focused on the \ce{CO2 + H <--> HCOO} reaction on \ce{In_yO_x/Cu(111)}.
The geometries of formate adsorbed on inverse catalysts from our earlier work~\cite{kempen2025,nielsen2025} form the basis for the training data in this work.
We included \ce{In_yO_x} nanoclusters with \num{3} to \num{6} indium atoms and \num{2} to \num{3} oxygen atoms, including all nanoclusters within \qty{0.1}{\electronvolt} of the global minimum energy.
From these formate-adsorbed nanoclusters and corresponding clean nanoclusters, structures approximating the reaction pathway were constructed using the following procedure:

\begin{enumerate}
    \itemsep0pt
    \item Sample formate-adsorbed structures from our earlier work~\cite{nielsen2025};
    \item For each formate-adsorbed structure, generate a corresponding \ce{CO2}-adsorbed structure by removing the hydrogen atom from formate and relaxing the resulting structure;
    \item Add an \ce{H} atom to the metal surface on the three-fold site closest to the \ce{CO2} adsorbate and relax the resulting structure;
    \item Generate an initial guess of a reaction path by interpolating \num{10} images between the initial state (\ce{CO2{*} + H{*}}) and the corresponding final state (\ce{HCOO{*}}) using an image dependent pair potential (IDPP) path~\cite{smidstrup2014} with a force convergence of $F_\mathrm{max} = \qty{0.1}{\electronvolt\per\angstrom}$, and perform single-point DFT evaluations for these interpolated images;
    \item With the highest-energy image from the interpolated path, carry out two short (i.e., not converged) local TS optimizations in parallel; one using the ClimbFixInternals methods from ASE~\cite{larsen2017,plessow2018} and one using the bond climbing method outlined in \cref{sec:methods/ts-search}.
\end{enumerate}

Using the above approach, we sampled \num{48} pairs of initial and final states of the \ce{CO2 + H <--> HCOO} reaction.
Structures from all the steps were combined into a single data set, where only a small subset of the structures from step 5 were chosen due to their high relative similarity.
This resulted in a total of \num{1120} structures.
Due to practical memory constraints, training the GM-NN models was only done on a randomly-selected subset of \num{694} structures; this was further split into training and validation sets using a $\num{90}:\num{10}$ split.
The remaining \num{426} structures were used as testing data.

The \num{48} pairs of initial and final states used for training data acquisition were also used for TS search (outlined below), forming an in-sample TS data set for searching purposes.
Additionally, steps 1 through 3 of the procedure above were repeated for \num{76} new pairs of initial and final states, ensuring that the sampled adsorption geometries are distinct from the ones included in the training dataset, to generate a fully out-of-sample TS data set.

\subsection{Transition state search}  \label{sec:methods/ts-search}

TS searches were performed on the in-sample TS data set consisting of \num{48} pairs of initial and final states of the \ce{CO2 + H <--> HCOO} reaction, as well as on the \num{76} pairs of initial and final states forming the out-of-sample TS data set.
Using the set of 10 independently trained GM-NN models (see \cref{sec:methods/gmnn}), \num{10} TS searches were performed for each pair of initial and final states using a two-step procedure.

The first step consists of a nudged elastic band (NEB) calculation~\cite{jonsson1998}.
To begin, \num{40} images are interpolated between the initial and the final state using the IDPP interpolation method with a force convergence of $F_\mathrm{max} = \qty{0.01}{\electronvolt\per\angstrom}$.
The NEB calculation is then performed with the IDPP path as a starting guess, and the minimum energy path is evolved using the \texttt{ode12r} solver~\cite{makri2019} until either convergence ($F_\mathrm{max} < \qty{0.05}{\electronvolt\per\angstrom}$) or stagnation of the solver.
The NEB force evaluation is switched to the climbing image method~\cite{henkelman2000} if $F_\mathrm{max} < \qty{0.10}{\electronvolt\per\angstrom}$.
The highest-energy image, if not yet optimized until convergence, is then selected for further refinement with a local bond climbing method~\cite{cheula_in_prep}.
In this method, the forces acting on the carbon and hydrogen atoms---between which a bond is to be formed in the formation of formate---are reflected with respect to the plane perpendicular to the bond vector.
This path is then similarly evolved using the \texttt{ode12r} solver until convergence or stagnation.
During TS search, only the nanocluster and adsorbate atoms are allowed to move.

Validation of TS guesses produced by the GM-NN model was performed by continuing the bond climbing method outlined above with DFT instead of with the GM-NN model until convergence.
This allows for energetic and geometric comparison between the GM-NN-converged TS structure and the DFT-converged TS structure.

\subsection{Active learning} \label{sec:methods/active-learning}

An active learning workflow was created to attempt to iteratively improve the GM-NN models.
The IPSuite framework~\cite{zills2024b,zills2024a} was utilized to set up the workflow, benefitting
from its native integration with Apax.

First, with the training and validation data sets outlined in \cref{sec:methods/training-data} and hyperparameters determined as outlined in \cref{sec:methods/gmnn}, a GM-NN model was trained.

Second, the model was used as a potential for performing NEB calculations for all \num{48} binding sites with initial--final pairs.
These calculations were initialized using IDPP-interpolated paths with \num{10} interpolated images and an IDPP force convergence of $F_\mathrm{max} = \qty{0.1}{\electronvolt\per\angstrom}$.
The minimum energy path was evolved using the \texttt{ode12r} solver either until convergence ($F_\mathrm{max} < \qty{0.05}{\electronvolt\per\angstrom}$) or until the uncertainty on the forces predicted by the shallow ensemble exceeded the magnitude of the force on any atom in the interpolated images.

Third, the \num{10} structures from the full NEB trajectories predicted to most improve model training were selected.
The selection was done by evaluating the gradient of the last-layer weights of the network with respect to each candidate structure~\cite{zaverkin2022}.
A gradient kernel was then established by considering all combinations of pairs of structures, and farthest point sampling (greedy distance maximization) was applied to this gradient kernel to select the structures.
This corresponds to the method referred to as \textsc{MaxDist} + FEAT(LL) by \citet{zaverkin2022}.

Fourth, the \num{10} selected structures were evaluated using single-point DFT calculations as outlined in \cref{sec:methods/dft} and added to the training data set.

After expanding the training data set and retraining the model, the workflow returns to the second step where NEB calculations are again initialized from IDPP-interpolated paths.
Every third iteration of the workflow, the GM-NN model is retrained from scratch in two steps: first an MSE model, then a shallow ensemble.
The other iterations, only the shallow ensemble model is retrained, initialized from the latest available MSE model.
This approach balances time spent on training with model prediction quality.

\section{Results}

\subsection{Model training}

\begin{figure}[t]
    \centering
    \includegraphics{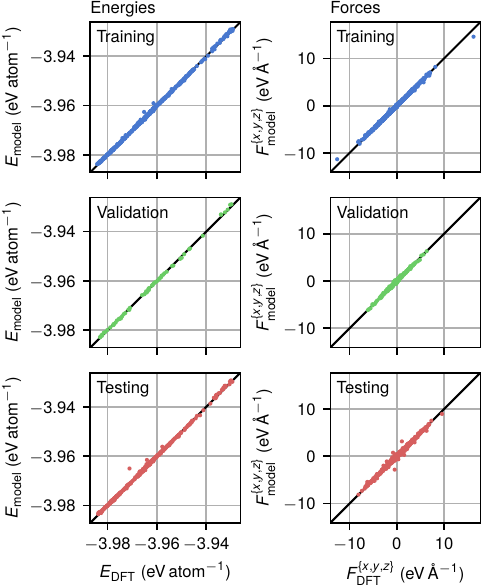}
    \caption{Parity plots showing prediction performance of one of the trained GM-NN models on the full training data. The force predictions shown include forces on the nanocluster and formate atoms.}
    \label{fig:model-training-parity}
\end{figure}

We begin by assessing the performance of the \num{10} separate GM-NN models all trained on the same initial training data set.
\cref{fig:model-training-parity} shows parity plots for one of the models, which achieves a root mean square error (RMSE) of \qty{0.43}{\milli\electronvolt\per\atom} for energy predictions and \qty{62}{\milli\electronvolt\per\angstrom} for force component predictions (on the nanocluster and formate atoms) on the testing set.
On average, all \num{10} models achieve an RMSE of \qty{0.54}{\milli\electronvolt\per\atom} for energies and \qty{69}{\milli\electronvolt\per\angstrom} for force components on the testing set.
RMSEs for all \num{10} models on all data subsets are provided in \cref{supp/sec:model-training}.
Due to stochastic model training, not all models perform equally well; two models in particular stand out for performing worse than the average.

\subsection{In-sample TS search}  \label{sec:results/ts-search}

With the \num{10} trained GM-NN models, we subsequently performed TS search on the in-sample set of \num{48} pairs of initial and final states of the \ce{CO2 + H <--> HCOO} reaction.
Since some of the models produced unreasonable TS guesses, we applied the following filtering steps.
First, a check of atomic distances was performed and structures with atoms too close to each other were filtered out.
The remaining structures were evaluated with a single-point DFT calculation to obtain accurate energy values.
Secondly, we discarded all structures that were more than \qty{1}{\electronvolt} higher in energy than the lowest-energy structure.
Finally, to validate the TS guesses, we randomly selected \num{20} out of the \num{48} reaction paths and continued the TS optimization by running the bond climbing method with DFT for all ML-optimized TS guesses remaining after the filtering step.

\begin{figure}[t]
    \centering
    \includegraphics{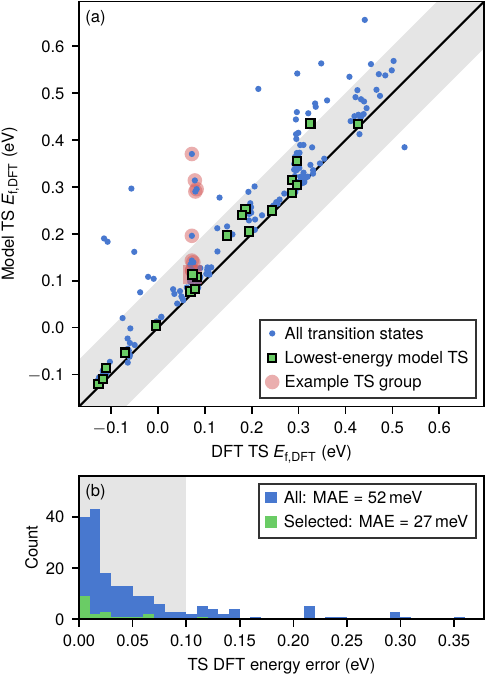}
    \caption{Comparison of energies of in-sample TS guesses produced by the GM-NN models to those of validated TSs produced by continuing bond climbing with DFT. (a) Parity plot of formation energies. The TS guesses highlighted in red correspond to one group of TSs from the same pair of initial and final states. (b) Histogram of errors. The highlighted band in both plots indicates an error below \qty{0.1}{\electronvolt}. The data points highlighted in green correspond to those selected using the selection method described in the text.}    \subfiglabel{fig:ts-parity-dft-parity}    \subfiglabel{fig:ts-parity-dft-error}    \label{fig:ts-parity-dft}
\end{figure}

In \cref{fig:ts-parity-dft}, we show how the formation energies associated with the in-sample TS guess structures produced by the GM-NN models compare to those of the validated TS structures obtained with DFT optimization.
\qty{84}{\percent} of the ML-optimized TS structures retained after filtering are within \qty{0.1}{\electronvolt} of the DFT-optimized TS structures and the average energy error is \qty{52}{\milli\electronvolt}.
Note that the energies for both the ML- and DFT-optimized TS structures (as well as for the initial and final state structures) are obtained from a single-point DFT evaluation using high-level settings, as described in \cref{sec:methods/dft}.
The TSs for which the error is large are generally also characterized by the DFT validation step displacing the reactant atoms by a relatively large amount, as shown in \cref{supp/fig:ts-validation-displacements}.

It should be noted that the trained GM-NN models achieve this accuracy mostly due to them having learned the general \emph{shape} of the potential energy surface well; the activation energies predicted by the GM-NN for the TS guesses can be quite far off the DFT-evaluated energies for the same geometries.
As shown in \cref{supp/fig:ts-parity-model}, the mean absolute error of GM-NN activation energy predictions for model-optimized TSs is about \qty{0.2}{\electronvolt}.
Therefore, a single-point DFT evaluation of the model-generated TS guess is still required.

Using the procedure described above, the achieved average energy error of \qty{52}{\milli\electronvolt} is already excellent and far below intrinsic errors of DFT with generalized gradient approximation functionals for surface adsorption energies, which are typically about \qtyrange{0.2}{0.3}{\electronvolt}~\cite{Wellendorff2015}.
Still, we investigated whether it would be possible to improve on either TS guess structures or energies by acquiring further training data in an active learning framework.
However, this did not turn out to meaningfully improve either model prediction performance or TS search performance.
We discuss these active learning results in more detail in \cref{supp/sec:active-learning}.

Nevertheless, our results highlight the power of ML-based TS optimization.
The GM-NN models have learned the potential energy surface well enough that TS geometries can be guessed with high accuracy, requiring only a single-point DFT evaluation to obtain the corresponding potential energy.
Continuing bond climbing with DFT, as done here for validation purposes, took a median of \num{106} single-point DFT evaluations per TS guess, for an average energy accuracy improvement of only \qty{52}{\milli\electronvolt}.

Interestingly, the DFT validation process as applied in this work tends to lower the TS energies, i.e., TS guesses produced by the GM-NN models are only rarely underpredicting the corresponding formation energy.
This is most likely due to the bond climbing method being equivalent to a regular relaxation for all non-reacting atoms, and the energy decrease from relaxing these atoms into their DFT local minimum dominates any energy increase from reacting atoms moving against the force direction.
As we have \num{10} independently-generated TS candidates for every pair of initial and final states (as exemplified with the highlighted points in \cref{fig:ts-parity-dft-parity}), we can use this observation to guide TS selection from this set of candidates.
Under the assumption that TS guesses are never underpredicted, the model-generated TS guess with the lowest energy must be closest in energy to a DFT-validated TS.
While this assumption is not always valid, it is valid often enough that selecting TSs with this criterion increases accuracy even further.
The TSs selected using this method are indicated in green in \cref{fig:ts-parity-dft}.
With this selection, the average energy error between the model-generated TSs and DFT-validated TSs is reduced from \qty{52}{\milli\electronvolt} to \qty{27}{\milli\electronvolt}, and all but one of the selected TSs have an error below \qty{0.1}{\electronvolt}.

\subsection{Out-of-sample TS search}

\begin{figure}[t]
    \centering
    \includegraphics{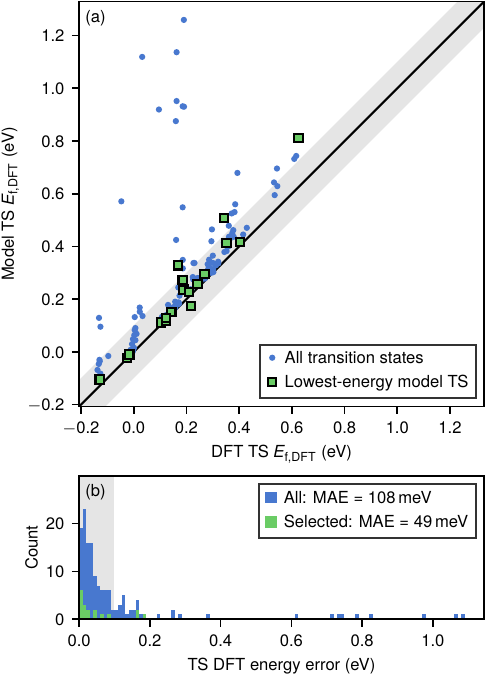}
    \caption{Comparison of energies of out-of-sample TS guesses produced by the GM-NN models to those of validated TSs obtained by continuing bond climbing with DFT. A detailed description is given in the caption of \cref{fig:ts-parity-dft}.}    \subfiglabel{fig:ts-parity-dft-out-of-sample-parity}    \subfiglabel{fig:ts-parity-dft-out-of-sample-error}    \label{fig:ts-parity-dft-out-of-sample}
\end{figure}

Encouraged by the excellent accuracy for the in-sample TS search, we next test the accuracy of our approach for the out-of-sample set of \num{76} new pairs of initial and final states, applying the same filtering procedure as for the in-sample data set, and continuing bond climbing with DFT for another random subset of \num{20} TSs.
The activation energies for these out-of-sample TS guesses are compared to the DFT-validated TSs in \cref{fig:ts-parity-dft-out-of-sample}.
For the out-of-sample data set, prediction is slightly worse than for the in-sample data set, with \qty{76}{\percent} of the TS guesses accurate within \qty{0.1}{\electronvolt} to the DFT-validated TSs.

Analogous to the in-sample data set, applying the TS selection criterion outlined in \cref{sec:results/ts-search} improves accuracy on the out-of-sample data set, reducing the average energy error from \qty{108}{\milli\electronvolt} across the full data set to \qty{49}{\milli\electronvolt} for the selected data set.
Direct activation energy prediction by the GM-NN is also similarly performant for the in-sample and out-of-sample data sets, with the latter having a mean absolute error of \qty{0.22}{\electronvolt} (see \cref{supp/fig:ts-parity-model-out-of-sample}).

Overall, these results highlight that the trained GM-NNs are not only performant on the pairs of initial and final states they have been trained on, but also on geometries not encountered during training.
This makes the outlined approach particularly suitable for screening a large data set of nanocluster and adsorbate geometries.

\subsection{TS structure--activity trends}

\begin{figure}[t]
    \centering
    \includegraphics{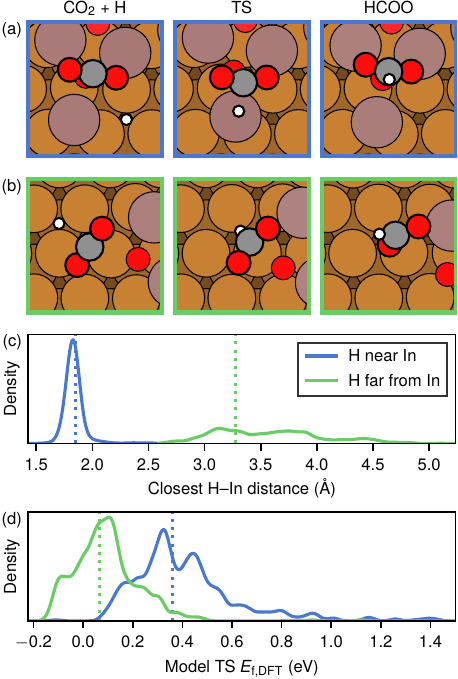}
    \caption{Two distinct classes of TSs. (a--b) Top-down visualizations of (left-to-right) initial state, TS, and final state. (a) A TS where hydrogen is located near indium; (b) a TS where hydrogen is located far from indium. (c) Closest \ce{H}--\ce{In} distance distribution for all TSs, showing the distinction in two classes. (d) Formation energy distributions for these two classes of TSs. The dotted lines in (c--d) indicate the distances and formation energies for the examples shown in (a--b).}    \subfiglabel{fig:ts-near-indium}    \subfiglabel{fig:ts-far-indium}    \subfiglabel{fig:ts-near-far-indium-distance}    \subfiglabel{fig:ts-near-far-indium-eact}    \label{fig:ts-near-far-indium}
\end{figure}

With the extensive set of TSs identified, made possible by our computationally efficient workflow relying on TS optimization in the GM-NN model, we next analyze structure--activity trends.
For this we employ the combined set of in-sample and out-of-sample model-optimized TS structures, selecting always the TS structure with the lowest DFT energy for each pair of initial and final state, leading to a total number of 124 TSs.
\cref{fig:ts-near-far-indium} shows that the TSs discovered in this work can be categorized into two distinct classes based on the closest \ce{H}--\ce{In} distance in the TS.
\cref{fig:ts-near-far-indium-distance} clearly shows a bimodal distribution with a relatively narrow peak for \ce{H} near \ce{In}, and a much broader peak for \ce{H} further away from \ce{In}.
The case where \ce{H} is near \ce{In} describes TSs where the hydrogen atom reacts with the \ce{CO2} adsorbate \emph{via} a nanocluster indium atom, as shown in \cref{fig:ts-near-indium}.
The case where \ce{H} is far from \ce{In} describes TSs where the hydrogen atom more directly binds to \ce{CO2} from the metal surface, as exemplified in \cref{fig:ts-far-indium}.
The preference for either of these classes is mostly dictated by geometry; indium atoms present near the hydrogen atom and the \ce{CO2} compound can both offer a site \emph{via} which hydrogen can react, but also hinder the hydrogen atom from approaching \ce{CO2} directly.

The formation energy distributions for these two cases are shown in \cref{fig:ts-near-far-indium-eact}.
The division in \ce{H}--\ce{In} distance somewhat correlates with formation energies; those where \ce{H} is transferred \emph{via} \ce{In} are on average higher than those for reactions where this does not occur.
This is explained by the fact that the adsorption of \ce{H} on an \ce{In} nanocluster atom is approximately \qty{0.4}{\electronvolt} less stable than that of \ce{H} on the \ce{Cu} surface.

\subsection{Scaling relations}

\begin{figure}[t]
    \centering
    \includegraphics{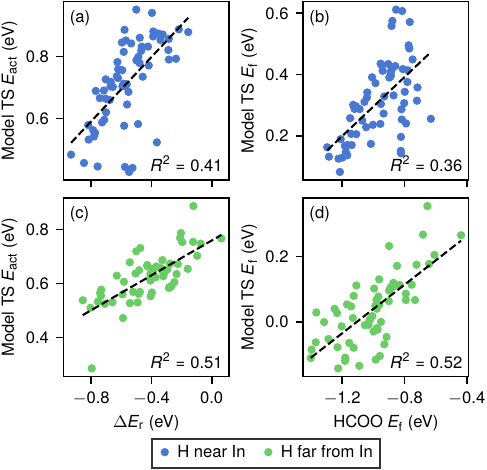}
    \caption{Two types of scaling relations for the formate formation step, distinguishing the two separate TS classes discussed in the text. (a, c) BEP relation; (b, d) relationship between formation energies of the TS and of formate.}
    \label{fig:scaling-relations}
\end{figure}

Linear scaling relations have in the past been found to be good, simple descriptors of catalytic processes on metal surfaces~\cite{norskov2002,Michaelides2003}.
The data collected in this work allows for validating whether these correlations also apply to the inverse catalysts explored here.
Two types of scaling relations have been evaluated (see \cref{fig:scaling-relations}).
The first is the BEP relation, relating activation energy $E_\mathrm{act}$ to reaction energy $\Delta E_\mathrm{r}$, the reference state being \ce{H} and \ce{CO2} adsorbed to the surface.
The second is a relation between the formation energy of the TS and that of formate, the reference state being \ce{H2} and \ce{CO2} in the gas phase.

Since we have found two classes of TSs in the previous section, we evaluate the scaling relations for these classes separately.
As visible, the linear relation best describing the data points is a quite poor predictor of the activation energy; prediction errors can be as large as half the full range of activation energies.
The relations are better for the class of TSs where hydrogen is not near indium.
This is likely because the geometries of these TSs are more similar to that of the final state, formate, than the geometries of TSs where hydrogen reacts \emph{via} an indium atom.
The poor fits observed underscore the need for more sophisticated methods of determining activation energies of reaction pathways on these more complex catalyst surfaces.

\section{Conclusions}

We have developed a ML-driven workflow for exploring energy barriers of elementary reaction steps at inverse catalysts, which captures the complexity associated with the possible cluster sizes and compositions of these catalysts.
By acquiring a suitable training data set consisting mainly of interpolated structures along the reaction pathway of the \ce{CO2 + H <--> HCOO} step at a diverse subset of cluster geometries, we show that it is possible to train a MLIP (a GM-NN model) that gives accurate TS guess structures for both in-sample cluster sites considered in the training data as well as out-of-sample sites.
While model-predicted activation energies based on these structures lack predictive accuracy with a MAE of approximately \qty{0.2}{\electronvolt}, a single-point DFT evaluation of the TS structure guess brings down the TS energy error to a MAE of \qty{52}{\milli\electronvolt} for in-sample TSs.
Furthermore, we introduce a TS guess structure selection method that brings down the error even further:
Realizing that TS guess structures predicted by independently trained GM-NN models all tend to be higher in energy than the DFT-optimized TS structure, selecting the model TS structure guess with the lowest DFT energy further lowers the MAE to \qty{27}{\milli\electronvolt}.
This procedure is also performant for out-of-sample cluster sites, where we achieve an only slightly higher MAE of \qty{49}{\milli\electronvolt}.

Finally, we make use of the combined data set of \num{124} model-optimized TS structures to analyze structure--activity trends.
We find that the TS structures fall into two categories depending on whether migration of the \ce{H} atom occurs over the \ce{Cu} surface next to the cluster edge or over an \ce{In} atom, where the former structures tend to be lower in energies due to favorable \ce{Cu}--\ce{H} interactions.
Scaling relations are obeyed for none of the two TS structure categories.
In fact, we observe several TS structures that exhibit catalytically interesting deviations from scaling relations, i.e., it is possible to have a low TS formation energy despite a moderate formate formation energy.
Such sites could make it possible to overcome limitations posed by the Sabatier principle (as quantified by scaling relation-based volcano plots in heterogeneous catalysis~\cite{Medford2015}) by allowing facile activation and adsorption of the reactants (i.e., \ce{CO2} and \ce{H2}) without poisoning the surface with too-strongly bound intermediates, which would tend to make product formation steps (e.g., methanol, methane, or carbon monoxide formation) difficult.

In this connection it is noteworthy that a recent experimental study has found that the \ce{In_yO_x/Au(111)} inverse catalyst is particularly active and selective for methanol formation, where the reaction pathway is expected to occur via the crucial formate intermediate~\cite{Prabhakar2024}.
Similarly, previous studies of other inverse catalyst systems such as \ce{Ce_yO_x/Cu(111)} and \ce{Zn_yO_x/Cu(111)} have highlighted their exceptional catalytic properties.
Based on our results, it seems likely that these catalytic properties could be related to the ability of the active site motifs found at inverse catalyst to break scaling relations.

\section*{Data availability}

The structure datasets will be made available on Zenodo upon publication under a CC BY 4.0 license.

\section*{Code availability}

The workflow scripts will be made available upon publication under a GNU GPLv3 license.

\section*{Supporting Information}

Additional information on model training, TS guess energies, and active learning (PDF)

\section*{Acknowledgments}

M.A. acknowledges funding from the European Union's Horizon 2020 research and innovation programme under the Marie Skłodowska-Curie grant agreement No 754513 and The Aarhus University Research Foundation, the Danish National Research Foundation through the Center of Excellence `InterCat' (grant no.\@ DNRF150), and VILLUM FONDEN (grant no.\@ 37381). Computational support was provided by the Centre for Scientific Computing Aarhus (CSCAA) at Aarhus University.

\section*{Author contributions}

\textbf{Luuk H. E. Kempen}: Conceptualization (equal); Investigation (lead); Methodology (lead); Software (lead); Visualization (lead), Writing -- original draft (lead); Writing -- review \& editing (equal).
\textbf{Marius Juul Nielsen}: Investigation (supporting); Methodology (supporting); Visualization (supporting); Writing -- review \& editing (equal).
\textbf{Mie Andersen}: Conceptualization (equal); Investigation (supporting); Writing -- review \& editing (equal); Supervision (lead); Resources (lead); Funding acquisition (lead).

\section*{Competing interests}

The authors declare no competing interests.

\printbibliography

\clearpage

\onecolumn
\raggedbottom

\setcounter{page}{1}
\setcounter{section}{0}
\setcounter{table}{0}
\setcounter{figure}{0}
\setcounter{equation}{0}
\renewcommand{\thepage}{S\arabic{page}}
\renewcommand{\thesection}{S\arabic{section}}
\renewcommand{\thetable}{S\arabic{table}}
\renewcommand{\thefigure}{S\arabic{figure}}
\renewcommand{\theequation}{S\arabic{equation}}

\makeatletter
\let\oldtitle\@title
\title{Supporting Information: ``\oldtitle''}
\date{}
\makeatother

\maketitle

\begin{refsection}

\section{Model training}  \label{supp/sec:model-training}

\begin{table}[H]
    \centering
    \caption{GM-NN model hyperparameters used for the models in this work. If no value is listed in the `shallow ensemble' column, the value is the same as for the single network.}
    \label{supp/tab:model-hyperparameters}
    \begin{tabular}{lcc}
        \toprule
        & \multicolumn{2}{c}{Value} \\
        \cmidrule(l){2-3}
        Hyperparameter & Single network & Shallow ensemble \\
        \midrule
        Radial basis function & Gaussian \\
        Number of Gaussians & \num{13} \\
        Basis function $R_\mathrm{min}$ & \qty{0.72}{\angstrom} \\
        Basis function $R_\mathrm{max}$ & \qty{7.0}{\angstrom} \\
        Number of contractions & \num{7} \\
        Number of basis functions & \num{11} \\
        Empirical correction & exponential \\
        Empirical correction $R_\mathrm{max}$ & \qty{1.22}{\angstrom} \\
        Number of hidden layers & \num{4} \\
        Number of nodes in hidden layers & \num{51}, \num{71}, \num{82}, \num{35} \\
        Embedding initialization & uniform in $[-1, 1]$ \\
        Weight initialization & LeCun~\cite{lecun1998} \\
        Bias initialization & \num{0} \\
        Optimizer & AdamW~\cite{loshchilov2019} \\
        Embedding contraction coefficient learning rate & \num{2.41e-3} & \num{2.41e-5} \\
        Neural network parameter learning rate & \num{1.17e-3} & \num{1.17e-5} \\
        Elemental output scaling factor learning rate & \num{1.03e-3} & \num{1.03e-5} \\
        Elemental output shift learning rate & \num{2.87e-3} & \num{3.87e-5} \\
        Empirical correction scaling factor learning rate & \num{0.507} & \num{0.01} \\
        Empirical correction prefactor learning rate & \num{0.444} & \num{0.01} \\
        Per-element gradient clipping & \num{5.53} \\
        Weight decay & \num{3.91e-4} \\
        Learning rate schedule & linear \\
        \bottomrule
    \end{tabular}
\end{table}

\begin{table}[H]
    \centering
    \caption{Model prediction RMSEs for all \num{10} models on the data generated for model training.}
    \label{supp/tab:model-predictions}
    
\begin{tabular}{cS[table-format=1.2]S[table-format=1.2]S[table-format=1.2]S[table-format=2]S[table-format=2]S[table-format=2]}
    \toprule
    & \multicolumn{3}{c}{Energies (\unit{\milli\electronvolt\per\atom})} & \multicolumn{3}{c}{Force components (\unit{\milli\electronvolt\per\angstrom})} \\
    \cmidrule(lr){2-4}
    \cmidrule(lr){5-7}
    Model index & {Training} & {Validation} & {Testing} & {Training} & {Validation} & {Testing} \\
    \midrule
    1 & 0.30 & 0.36 & 0.43 & 30. & 58. & 62.  \\
2 & 0.90 & 0.87 & 0.91 & 80. & 92. & 91.  \\
3 & 0.31 & 0.29 & 0.45 & 43. & 61. & 58.  \\
4 & 0.76 & 0.75 & 0.75 & 83. & 94. & 98.  \\
5 & 0.42 & 0.38 & 0.40 & 31. & 62. & 62.  \\
6 & 0.26 & 0.31 & 0.31 & 19. & 57. & 64.  \\
7 & 0.43 & 0.38 & 0.51 & 30. & 56. & 56.  \\
8 & 0.53 & 0.61 & 0.59 & 41. & 66. & 66.  \\
9 & 0.46 & 0.49 & 0.50 & 36. & 64. & 65.  \\
10 & 0.45 & 0.48 & 0.51 & 28. & 56. & 68.  \\
    \midrule
    Average & 0.48 & 0.49 & 0.54 & 42. & 67. & 69.  \\
    \bottomrule
\end{tabular}

\end{table}

\begin{figure}[H]
    \centering
    \includegraphics{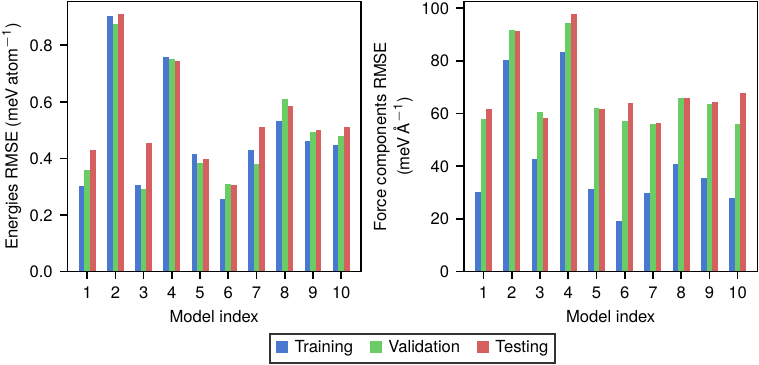}
    \caption{Model prediction RMSEs for all \num{10} models on the data generated for model training.}
    \label{supp/fig:model-predictions}
\end{figure}

\clearpage

\section{Transition state search}

\begin{figure}[H]
    \centering
    \includegraphics{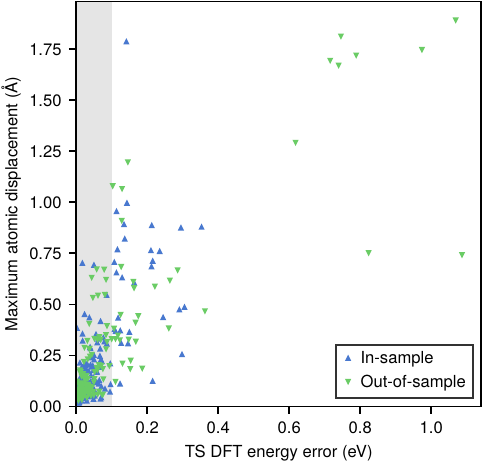}
    \caption{Correlation between DFT energy differences of TS guesses produced by the GM-NN models to those of validated TSs obtained by continuing bond climbing with DFT and the maximum displacement of reactant atoms as a result of continued bond climbing. The highlighted band indicates an error below \qty{0.1}{\electronvolt}.}
    \label{supp/fig:ts-validation-displacements}
\end{figure}

\begin{multicols}{2}

    \begin{figure}[H]
        \centering
        \includegraphics{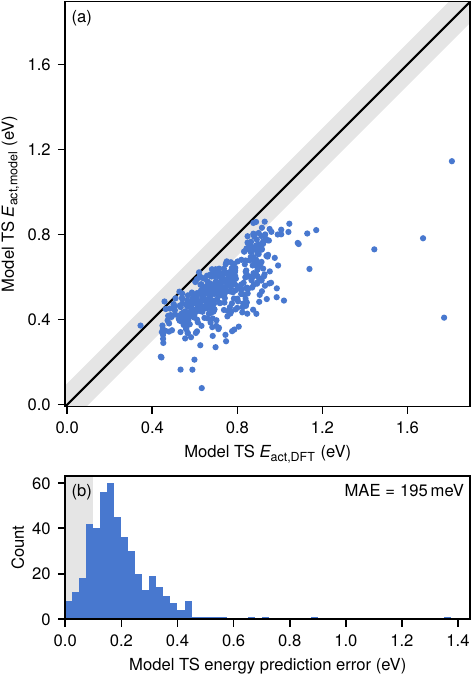}
        \caption{Comparison of model-predicted energies of in-sample TS guesses produced by the GM-NN models to DFT-evaluated energies of these geometries. (a) Parity plot of activation energies. (b) Histogram of prediction errors. The highlighted band in both plots indicates an error below \qty{0.1}{\electronvolt}.}
        \label{supp/fig:ts-parity-model}
    \end{figure}

    \columnbreak

    \begin{figure}[H]
        \centering
        \includegraphics{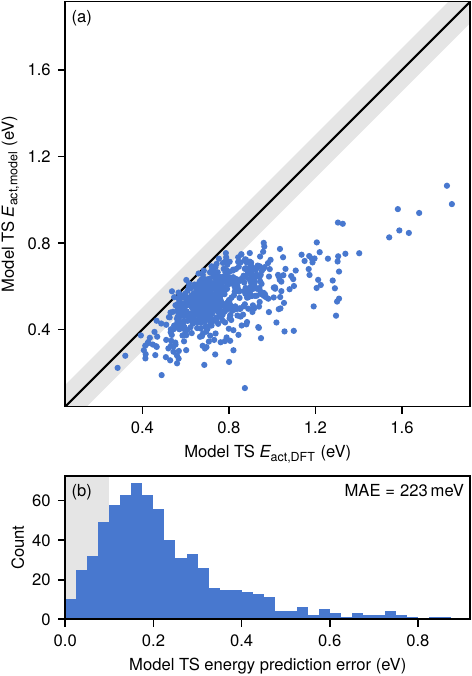}
        \caption{Comparison of model-predicted energies of out-of-sample TS guesses produced by the GM-NN models to DFT-evaluated energies of these geometries. (a) Parity plot of activation energies. (b) Histogram of prediction errors. The highlighted band in both plots indicates an error below \qty{0.1}{\electronvolt}.}
        \label{supp/fig:ts-parity-model-out-of-sample}
    \end{figure}

\end{multicols}

\section{Active learning}  \label{supp/sec:active-learning}

\begin{figure}[H]
    \centering
    \includegraphics{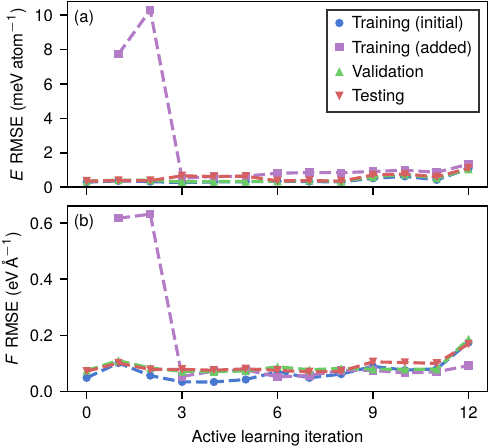}
    \caption{GM-NN model prediction performance in the active learning workflow. (a) Energies; (b) forces on the nanocluster and formate atoms.}
    \label{supp/fig:active-learning-rmses}
\end{figure}

\cref{supp/fig:active-learning-rmses} shows the prediction performance of the GM-NN models as data points sampled from NEB TS searches were added to the training set, as described in \cref{sec:methods/active-learning}.
Noteworthy is that these data points are far enough removed from the data present in the initial training set for them to be poorly predicted at first, even after retraining the shallow ensemble model.
This is evidenced by their significantly higher RMSE values at iterations \num{1} and \num{2} of the model, compared to those of the initial sets.
At iteration \num{3}, full retraining of the model from scratch did successfully incorporate this new class of data points into the model, and subsequent data iteratively added continued to be predicted with similar quality.

\begin{figure}[H]
        \centering
        \includegraphics{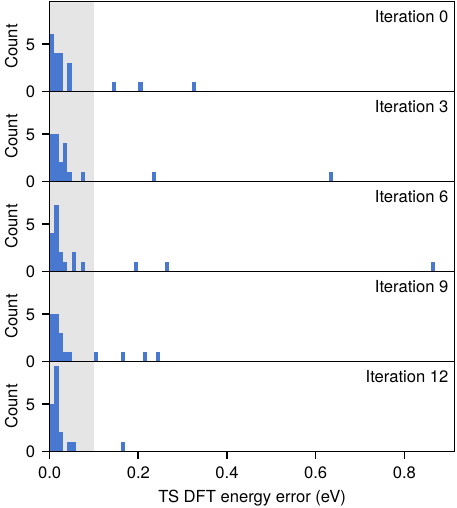}
        \caption{Comparison of energies of TS guesses produced by the iteratively trained GM-NN models to those of validated TSs by continuing bond climbing with DFT. The highlighted band indicates an error below \qty{0.1}{\electronvolt}.}
        \label{supp/fig:active-learning-ts-parity-dft}
    \end{figure}

\begin{figure}[H]
        \centering
        \includegraphics{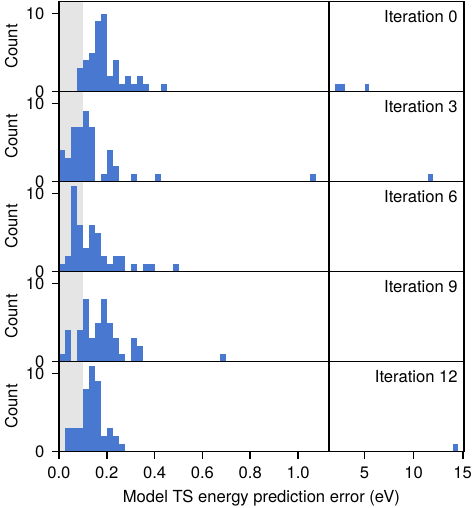}
        \caption{Comparison of model-predicted energies of TS guesses produced by the iteratively trained GM-NN models in the active learning workflow to DFT-evaluated energies of the same geometries. The highlighted band indicates an error below \qty{0.1}{\electronvolt}.}
        \label{supp/fig:active-learning-ts-parity-model}
    \end{figure}

More importantly, the active learning framework did not lead to any improvement in the quality of the guess TS structures.
In \cref{supp/fig:active-learning-ts-parity-dft}, model-generated TS guesses are compared with DFT-validated ones, and the average energy error stays relatively constant throughout active learning iterations.
Continuing bond climbing with DFT also does not become notably faster throughout the iterations.
Similarly, direct prediction of activation energies by the GM-NN also does not greatly improve during active learning, as shown in \cref{supp/fig:active-learning-ts-parity-model}.
The reason for the ineffectiveness of the employed active learning strategy is currently unclear and requires further investigation.

\clearpage

\begin{multicols}{2}

    \columnbreak

\end{multicols}

\printbibliography

\end{refsection}

\end{document}